\documentclass{jpsj-suppl}
\usepackage{txfonts} 
\usepackage{epsfig}
\usepackage{graphicx}
\usepackage{slashed}
\title{Pion Photo- and Electroproduction and the Chiral MAID Interface}

\author{Marius \textsc{Hilt}$^{1}$, Bj\"orn C.~\textsc{Lehnhart}$^{1}$, Stefan \textsc{Scherer}$^{1}$, and Lothar \textsc{Tiator}$^{1}$}

\inst{$^{1}$PRISMA Cluster of Excellence, Institut f\"ur Kernphysik,
Johannes Gutenberg-Universit\"at Mainz, D-55099 Mainz, Germany}

\email{scherer@kph.uni-mainz.de}

\recdate{October 16, 2015}

\abst{
   We discuss the extended on-mass-shell scheme for
manifestly Lorentz-invariant baryon chiral perturbation theory.
   We present a calculation of pion photo- and electroproduction
up to and including order $q^4$.
   The low-energy constants have been fixed by fitting experimental
data in all available reaction channels.
   Our results can be accessed via a web interface, the so-called chiral MAID
(http://www.kph.uni-mainz.de/MAID/chiralmaid/).}

\kword{chiral perturbation theory, pion photoproduction, pion
electroproduction}

\begin{document}
\maketitle

\section{Introduction}
\label{sec:introduction}
   In the middle of the 1980s, renewed interest in neutral pion photoproduction at threshold was triggered
by experimental data from Saclay and Mainz~\cite{Mazzucato:1986dz,Beck:1990da}, which indicated a serious
disagreement with the predictions for the $s$-wave electric dipole amplitude $E_{0+}$ based on current algebra
and PCAC \cite{DeBaenst:1971hp,Vainshtein:1972ih,Scherer:1991cy}.
   This discrepancy was explained with the aid of chiral perturbation theory (ChPT)~\cite{Bernard:1991rt}.
   Pion loops, which are beyond the current-algebra framework,
generate infrared singularities in the scattering amplitude which
then modify the predicted low-energy expansion of $E_{0+}$
(see also Ref.\ \cite{Davidson:1993et}).
   Subsequently, several experiments investigating pion photo- and
electroproduction in the threshold region were performed at Mainz,
MIT-Bates, NIKHEF, Saskatoon and TRIUMF, and on the theoretical side, all of
the different reaction channels of pion photo- and electroproduction
near threshold were extensively investigated by Bernard, Kaiser, and
Mei{\ss}ner within the framework of heavy-baryon chiral perturbation
theory (HBChPT) \cite{Bernard:1992qa}.
   For a complete list of references, see Ref.~\cite{Hilt:2013coa}.
   In the beginning, the manifestly Lorentz-invariant or relativistic formulation of ChPT (RChPT) was
abandoned, as it seemingly had a problem with respect to power counting when loops containing internal nucleon
lines come into play.
   In the meantime, the development of the infrared regularization (IR) scheme \cite{Becher:1999he}
and the extended on-mass-shell (EOMS) scheme \cite{Gegelia:1999gf,Fuchs:2003qc} offered a solution to
the power-counting problem, and RChPT became popular again.

   Here, we give a short introduction to the EOMS scheme and
present its application to a calculation of pion photo- and electroproduction
up to and including order $q^4$ [${\cal O}(q^4)$].
   We present the so-called chiral MAID ($\chi$MAID) \cite{website}
which provides the numerical results of these calculations.

\section{Renormalization and Power Counting}
   ChPT is the effective field theory of QCD in the low-energy regime \cite{Weinberg:1978kz,Gasser:1983yg,Gasser:1987rb}
(for an introduction, see Refs.~\cite{Scherer:2002tk,Scherer:2012zzd}).
   The prerequisite for an effective field theory program
is (a) a knowledge of the most general effective Lagrangian and (b) an expansion scheme for observables in terms
of a consistent power counting method.

\subsection{Effective Lagrangian and Power Counting}

  The effective Lagrangian relevant to the one-nucleon sector
consists of the sum of the purely mesonic and $\pi N$ Lagrangians, respectively,
\begin{displaymath}
{\cal L}_{\rm eff}={\cal L}_{\pi}+{\cal L}_{\pi N}={\cal L}_\pi^{(2)}+{\cal L}_\pi^{(4)}+\ldots+{\cal L}_{\pi
N}^{(1)}+{\cal L}_{\pi N}^{(2)}+\ldots,
\end{displaymath}
which are organized in a derivative and quark-mass expansion \cite{Weinberg:1978kz,Gasser:1983yg,Gasser:1987rb}.
   For example, the lowest-order basic Lagrangian ${\cal L}_{\pi N}^{(1)}$,
already expressed in terms of renormalized parameters and fields, is given by
\begin{equation}
\label{LpiN1} {\cal L}_{\pi N}^{(1)}=\bar \Psi \left( i\gamma_\mu
\partial^\mu - m\right) \Psi
-\frac{1}{2}\frac{\texttt{g}_A}{F} \bar \Psi \gamma_\mu \gamma_5 \tau^a \partial^\mu \pi^a \Psi +\ldots,
\end{equation}
where $m$, $\texttt{g}_A$, and $F$ denote the chiral limit of the physical nucleon mass, the axial-vector
coupling constant, and the pion-decay constant, respectively.
   The ellipsis refers to terms containing external fields and
higher powers of the pion fields.
   When studying higher orders in perturbation theory, one encounters
ultraviolet divergences.
   As a preliminary step, the loop integrals are regularized,
typically by means of dimensional regularization.
   In the process of renormalization the counter terms are adjusted such that
they absorb all the ultraviolet divergences occurring in the calculation of loop diagrams \cite{Collins:xc}.
   This will be possible, because we include in the Lagrangian all
of the infinite number of interactions allowed by symmetries \cite{Weinberg:1995mt}.
   When renormalizing, we still have the freedom of choosing a renormalization
prescription.
   In this context the finite pieces of the renormalized couplings  will be adjusted such that
renormalized diagrams satisfy the following power counting:
   a loop integration in $n$ dimensions counts as $q^n$,
pion and nucleon propagators count as $q^{-2}$ and $q^{-1}$, respectively, vertices derived from ${\cal
L}_{\pi}^{(2k)}$ and ${\cal L}_{\pi N}^{(k)}$ count as $q^{2k}$ and $q^k$, respectively.
   Here, $q$ collectively stands for a small quantity such as the pion
   mass, small external four-momenta of the pion, and small external
three-momenta of the nucleon.
   The power counting does not uniquely fix the renormalization scheme,
i.e., there are different renormalization schemes such as the IR \cite{Becher:1999he}
and EOMS \cite{Gegelia:1999gf,Fuchs:2003qc} schemes, leading to the above
specified power counting.

\subsection{Example: One-Loop Contribution to the Nucleon Mass}

\begin{figure}[t]
\begin{center}
\includegraphics[width=0.5\textwidth]{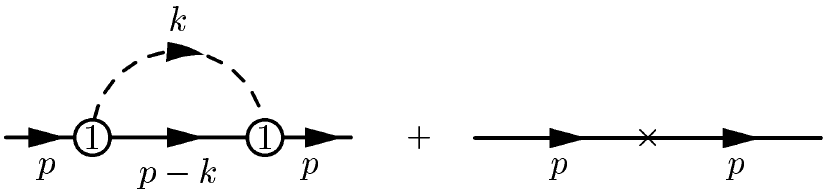}
\caption{\label{4:2:3:ren_diag} Renormalized one-loop self-energy
diagram. The number 1 in the interaction blobs refers to ${\cal L}_{\pi N}^{(1)}$.
The cross generically denotes counter-term contributions.
}
\end{center}
\end{figure}

   In the mesonic sector, the combination of dimensional regularization and
the modified minimal subtraction scheme $\widetilde{\mbox{MS}}$ leads to a straightforward correspondence between
the chiral and loop expansions \cite{Gasser:1983yg}.
   By discussing the one-loop contribution of Fig.~\ref{4:2:3:ren_diag}
to the nucleon self energy, we will see that this correspondence, at first sight, seems to be lost in the
baryonic sector.
   According to the power counting specified above, after renormalization,
we would like to have the order $D=n\cdot 1-2\cdot 1-1\cdot 1+1\cdot 2=n-1.$
   An explicit calculation yields \cite{Scherer:2012zzd}
\begin{displaymath}
\Sigma_{\rm loop}= -\frac{3 \texttt{g}_{A}^2}{4 F^2}\left\{
(\slashed{p}+m)I_N+M^2(\slashed{p}+m)I_{N\pi}-\frac{(p^2-m^2)\slashed{p}}{2p^2}[
(p^2-m^2+M^2)I_{N\pi}+I_N-I_\pi]\right\},
\end{displaymath}
where $M^2=2B\hat m$ is the lowest-order expression for the squared pion mass in terms of the low-energy coupling
constant $B$ and the average light-quark mass $\hat m$ \cite{Gasser:1983yg}.
   The relevant loop integrals are defined as
\begin{eqnarray}
\label{IpiIN}
I_\pi&=& \mu^{4-n}\int\frac{\mbox{d}^nk}{(2\pi)^n}\frac{i}{k^2-M^2+i0^+},\quad\quad
I_N=\mu^{4-n}\int\frac{\mbox{d}^nk}{(2\pi)^n}\frac{i}{k^2-m^2+i0^+},\\
\label{INpi} I_{N\pi}&=&\mu^{4-n}\int\frac{\mbox{d}^nk}{(2\pi)^n}
\frac{i}{[(k-p)^2-m^2+i0^+]}\frac{1}{k^2-M^2+i0^+}.
\end{eqnarray}
   The application of the $\widetilde{\rm MS}$ renormalization scheme of ChPT
\cite{Gasser:1983yg,Gasser:1987rb}---indicated by ``r''---yields
\begin{displaymath}
\Sigma_{\rm loop}^r=-\frac{3 \texttt{g}_{Ar}^2}{4 F^2_r} \left[M^2(\slashed{p}+m)I_{N\pi}^r+\ldots\right].
\end{displaymath}
   The expansion of $I_{N\pi}^r$ is given by
\begin{displaymath}
\label{4:2:3:INpExp}
    I_{N\pi}^r=\frac{1}{16\pi^2}\left(-1+\frac{\pi M}{m}+\ldots\right),
\end{displaymath}
resulting in $\Sigma_{\rm loop}^r={\cal O}(q^2)$.
   In other words, the $\widetilde{\rm MS}$-renormalized result does not
produce the desired low-energy behavior which, for a long time, was interpreted as the absence of a systematic
power counting in the relativistic formulation of ChPT.

   The expression for the nucleon mass $m_N$ is obtained by solving
the equation
\begin{displaymath}
\label{4:2:3:MassDef} m_N-m-\Sigma(m_N)=0,
\end{displaymath}
from which we obtain for the nucleon mass in the $\widetilde{\rm MS}$ scheme \cite{Gasser:1987rb},
\begin{equation}
\label{4:2:3:MassMStilde}
    m_N=m-4c_{1r}M^2+
    \frac{3\texttt{g}_{Ar}^2M^2}{32\pi^2F^2_r}m
    -\frac{3\texttt{g}_{Ar}^2M^3}{32\pi F^2_r}.
\end{equation}
   At ${\cal O}(q^2)$, Eq.~(\ref{4:2:3:MassMStilde}) contains, besides the undesired loop
contribution proportional to $M^2$, the tree-level contribution $-4c_{1r}M^2$ from the next-to-leading-order
Lagrangian ${\cal L}_{\pi N}^{(2)}$.

   The solution to the power-counting problem is the observation
that the term violating the power counting, namely, the third on the right-hand side of
Eq.~(\ref{4:2:3:MassMStilde}), is \emph{analytic} in the quark mass and can thus be absorbed in counter terms.
   In addition to the $\widetilde{\rm MS}$ scheme we have to perform an additional
{\em finite} renormalization.
   For that purpose we rewrite
\begin{equation}
\label{4:2:3:cRenorm}
    c_{1r}=c_1+\delta c_1,\quad \delta c_1 =\frac{3 m {\texttt g}_A^2}{128 \pi^2 F^2_r}+\ldots
\end{equation}
in Eq.~(\ref{4:2:3:MassMStilde}) which then gives the final result for the nucleon mass at ${\cal O}(q^3)$:
\begin{equation}
\label{4:2:3:MassFinal}
    m_N=m-4c_{1}M^2
    -\frac{3\texttt{g}_{A}^2M^3}{32\pi F^2}.
\end{equation}
   We have thus seen that the validity of a power-counting scheme is intimately
connected with a suitable renormalization condition.
   In the case of the nucleon mass, the $\widetilde{\rm MS}$ scheme alone does not
suffice to bring about a consistent power counting.

\subsection{Extended On-Mass-Shell Scheme}
   We illustrate the underlying ideas of the EOMS scheme in terms of a typical one-loop integral
in the chiral limit,
\begin{displaymath}
H(p^2,m^2;n)= -i\int \frac{{\mbox d}^n k}{(2\pi)^n} \frac{1}{[(k-p)^2-m^2+i0^+][k^2+i0^+]},
\end{displaymath}
where $\Delta=(p^2-m^2)/m^2={\cal O}(q)$ is a small quantity.
   Applying the dimensional counting analysis of
Ref.~\cite{Gegelia:1994zz}, the result of the integration is of the form
\begin{displaymath}
H\sim F(n,\Delta)+\Delta^{n-3}G(n,\Delta),
\end{displaymath}
where $F$ and $G$ are hypergeometric functions which are analytic for $|\Delta|<1$ for any $n$.
   The central idea of the EOMS scheme \cite{Gegelia:1999gf,Fuchs:2003qc} consists of
subtracting those terms which violate the power counting as $n\to 4$.
   Since the terms violating the power counting are analytic in small
quantities, they can be absorbed by counter-term contributions.
   In the present case, we want the renormalized integral to be of
the order $D=n-1-2=n-3$.
   To that end one first expands the integrand in
small quantities and subtracts those integrated terms whose order is smaller than suggested by the power
counting.
   The corresponding subtraction term reads \cite{Fuchs:2003qc}
\begin{displaymath}
H^{\rm subtr}=-i\int \frac{\mbox{d}^n k}{(2\pi)^n}\left. \frac{1}{[k^2-2p\cdot k +i0^+][k^2+i0^+]}\right|_{p^2=m^2}
=\frac{m^{n-4}}{(4\pi)^\frac{n}{2}}\frac{\Gamma\left(2-\frac{n}{2}\right)}{n-3},
\end{displaymath}
and the renormalized integral is written as
\begin{displaymath}
H^R=H-H^{\rm subtr}=\frac{m^{n-4}}{(4\pi)^{\frac{n}{2}}}\left[-\Delta\ln(-\Delta)+(-\Delta)^2\ln(-\Delta)+\ldots\right]
={\cal O}(q)\quad\text{as $n\to 4$.}
\end{displaymath}

\section{Pion Photo- and Electroproduction}
\label{sec:electroproduction}

\subsection{Invariant Amplitude and Cross Section}
   In the one-photon-exchange approximation, the invariant amplitude
for the reaction $e(k_i)+N(p_i)\rightarrow e(k_f)+N(p_f)+\pi(q)$ may be written as
\begin{displaymath}
{\cal M}=\epsilon_\mu {\cal M}^\mu,\quad \epsilon_\mu=e\frac{\bar{u}(k_f)\gamma_\mu u(k_i)}{k^2},\quad k=k_i-k_f,
\end{displaymath}
where $\epsilon_\mu$ denotes the polarization vector of the virtual photon
and ${\cal M}^\mu$ is the transition current matrix element:
${\cal M}^\mu=-ie\langle N(p_f),\pi(q)|J^\mu(0)|N(p_i)\rangle$.
   Using current conservation, $k_\mu{\cal M}^\mu=0$,
the transition current matrix element may be parameterized in terms of six invariant amplitudes $A_i$,
\begin{equation}
\label{eq:dennery}
\mathcal{M}^\mu=\bar{u}(p_f)\Bigg(\sum_{i=1}^6 A_i(s,t,u) M_i^\mu\Bigg)u(p_i),
\end{equation}
where the Mandelstam variables $s$, $t$, and $u$ satisfy
$s+t+u=2m_N^2+M_\pi^2-Q^2$ with $Q^2=-k^2$.
   The $M_i^\mu$ are suitable, linearly independent $4\times 4$ matrices such as, e.g.,
\begin{displaymath}
M_1^\mu=-\frac{i}{2}\gamma_5\left(\gamma^\mu \slashed{k}-\slashed{k}\gamma^\mu\right),\quad\ldots
\end{displaymath}
    For the purpose of performing a multipole expansion, we express the
invariant amplitude in the center-of-mass (cm) frame as \cite{Chew:1957tf,Dennery:1961zz}
\begin{displaymath}
{\cal M}=
\frac{4\pi W}{m_N}\chi_f^\dagger\mathcal{F}\chi_i,\quad\text{$\chi$: Pauli spinor,}
\end{displaymath}
with the Chew-Goldberger-Low-Nambu (CGLN) amplitudes $\mathcal{F}_i$ $(i=1,\ldots,6$),
\begin{align*}
\mathcal{F}=&i \vec{\sigma}\cdot\vec{a}_\perp \mathcal{F}_1(W,\Theta_\pi,Q^2)+\ldots.
\end{align*}
   The CGLN amplitudes can be expanded in a multipole series,
\begin{displaymath}
\mathcal{F}_1=\sum_{l=0}^\infty\Big\{\big[lM_{l+}+E_{l+}\big]P'_{l+1}(x)
+\big[(l+1)M_{l-}+E_{l-}\big]P'_{l-1}(x)\Big\},\quad \ldots
\end{displaymath}
where $x=\cos\Theta_\pi=\hat{q}\cdot\hat{k}$.
   Here, $P_l(x)$ is a Legendre polynomial of degree $l$,
$P'_l=dP_l/dx$ and so on, with $l$ denoting the orbital angular momentum
of the pion-nucleon system in the final state.
   The multipoles $E_{l\pm}$, $M_{l\pm}$, and $L_{l\pm}$ are functions of the
cm total energy $W$ and the photon virtuality $Q^2$ and
refer to transversal electric and
magnetic transitions and longitudinal transitions, respectively.
   The subscript $l\pm$ denotes the total angular momentum $j=l\pm1/2$ in the final state.
   Finally, in the isospin-symmetric limit, the four physical channels can be
expressed in terms of three isospin amplitudes (0), (+), and ($-$):
\begin{align*}
A_i(\gamma^{(*)}p\rightarrow n\pi^+)&=\sqrt{2}\left(A_i^{(-)}+A_i^{(0)}\right),&
A_i(\gamma^{(*)}n\rightarrow p\pi^-)&=-\sqrt{2}\left(A_i^{(-)}-A_i^{(0)}\right),\\
A_i(\gamma^{(*)}p\rightarrow p\pi^0)&=A_i^{(+)}+A_i^{(0)},
&A_i(\gamma^{(*)}n\rightarrow n\pi^0)&=A_i^{(+)}-A_i^{(0)}.
\end{align*}

    For pion photoproduction with polarized photons from an unpolarized target without recoil
polarization detection, the cross section can be written in the following way with the unpolarized
cross section $\sigma_0$ und the photon beam asymmetry $\Sigma$.
\begin{equation}
\frac{d \sigma}{d \Omega} =  \sigma_0 \left( 1 - P_T \Sigma \cos 2 \varphi \right).
\end{equation}
For $\pi^0$ photoproduction on the proton, both observables are very precisely measured in the
threshold region, allowing for an almost model independent partial wave
analysis~\cite{Hornidge:2012ca}.

   For pion electroproduction, in the one-photon-exchange approximation, the differential cross
section can be written as
\begin{equation}
\frac{d\sigma}{d\mathcal{E}_fd\Omega_fd\Omega_\pi^\textnormal{\,cm}}
=\Gamma\frac{d\sigma_v}{d\Omega_\pi^\textnormal{\,cm}},
\end{equation}
where $\Gamma$ is the virtual photon flux and $d\sigma_v/d\Omega_\pi^\textnormal{\,cm}$ is the pion
production cross section for virtual photons.

   For an unpolarized target and without recoil polarization detection, the virtual-photon
differential cross section for pion production can be further decomposed as
\begin{equation}
\frac{d\sigma_v}{d\Omega_\pi}=\frac{d\sigma_T}{d\Omega_\pi} +\epsilon\frac{d\sigma_L}{d\Omega_\pi}
+\sqrt{2\epsilon(1+\epsilon)}\frac{d\sigma_{LT}}{d\Omega_\pi}\cos\Phi_\pi
+\epsilon\frac{d\sigma_{TT}}{d\Omega_\pi}\cos{2\Phi_\pi}
+h\sqrt{2\epsilon(1-\epsilon)}\frac{d\sigma_{LT'}}{d\Omega_\pi}\sin{\Phi_\pi}, \label{eqn:wqparts}
\end{equation}
where it is understood that the variables of the individual virtual-photon cross sections
$d\sigma_T/d\Omega_\pi$ etc.~refer to the cm frame.
   For further details, especially concerning polarization observables, see Ref.~\cite{Drechsel:1992pn}.

\subsection{Evaluation of the Invariant Amplitude and Chiral MAID}
   At ${\cal O}(q^3)$, the invariant amplitude involves 15 tree-level diagrams and 50 one-loop diagrams.
   At ${\cal O}(q^4)$, 20 tree-level diagrams and 85 one-loop diagrams contribute.
   We have calculated the loop contributions numerically, using the computer algebra system MATHEMATICA with
the FeynCalc \cite{Mertig:1990an} and LoopTools packages \cite{Hahn:2000kx}.
   We have explicitly verified that current conservation and crossing symmetry
are fulfilled analytically for our results.

   At ${\cal O}(q^3)$, four independent LECs exist which are specifically related
to pion photoproduction.
   Two of them enter the isospin ($-$) channel and are, therefore, only relevant
for the production of charged pions.
   Moreover, they contribute differently to the invariant amplitudes $A_i$ of Eq.~(\ref{eq:dennery}).
   The remaining two constants enter the isospin
(+) and (0) channels, respectively, though both in combination with the same Dirac structure.
   Finally, at ${\cal O}(q^3)$ the description of pion electroproduction is a prediction,
because no new parameter (LEC) beyond photoproduction is available at that order.
   At ${\cal O}(q^4)$,  15 additional LECs appear.
   In the case of pion photoproduction, five constants contribute to the isospin (0) channel,
five constants to the isospin (+) channel, and one constant to the isospin ($-$) channel.
   For electroproduction, the (0) and (+) channels each have two more independent LECs.
   We note that the isospin $(-)$ channel, even at ${\cal O}(q^4)$, does not contain any
free LEC specifically related to electroproduction.

\begin{figure}[ht]
\includegraphics[width=\textwidth]{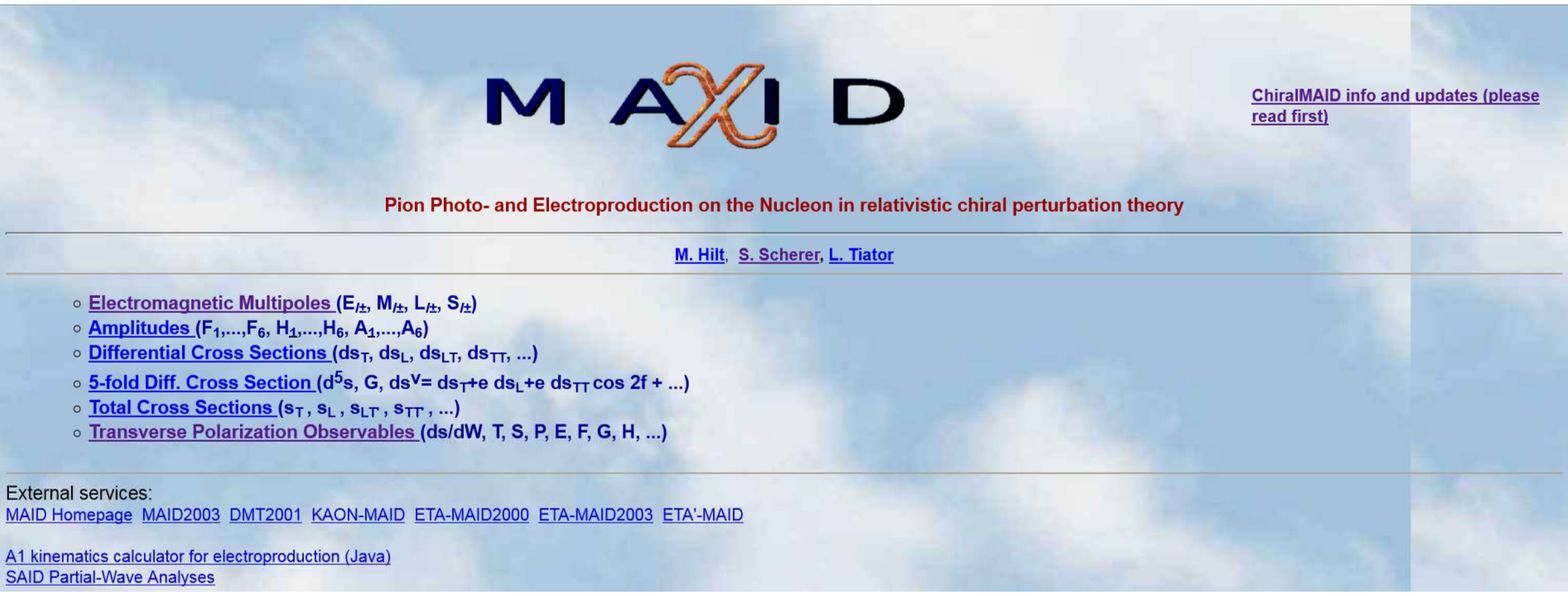}
\caption{\label{fig:chiral_MAID} Chiral MAID homepage [http://www.kph.uni-mainz.de/MAID/chiralmaid/].}
\end{figure}

   Figure \ref{fig:chiral_MAID} shows the homepage of the web interface chiral MAID.
   The loop contributions, including their parameters, are fixed and cannot be modified from the outside.
   On the other hand, the contact diagrams at ${\cal O}(q^3)$ and ${\cal O}(q^4)$ enter analytically and the
corresponding LECs can be changed arbitrarily (see Ref.~\cite{Hilt:2013coa} for a discussion of our present values).

\begin{figure}[ht]
\begin{center}
\includegraphics[width=\textwidth]{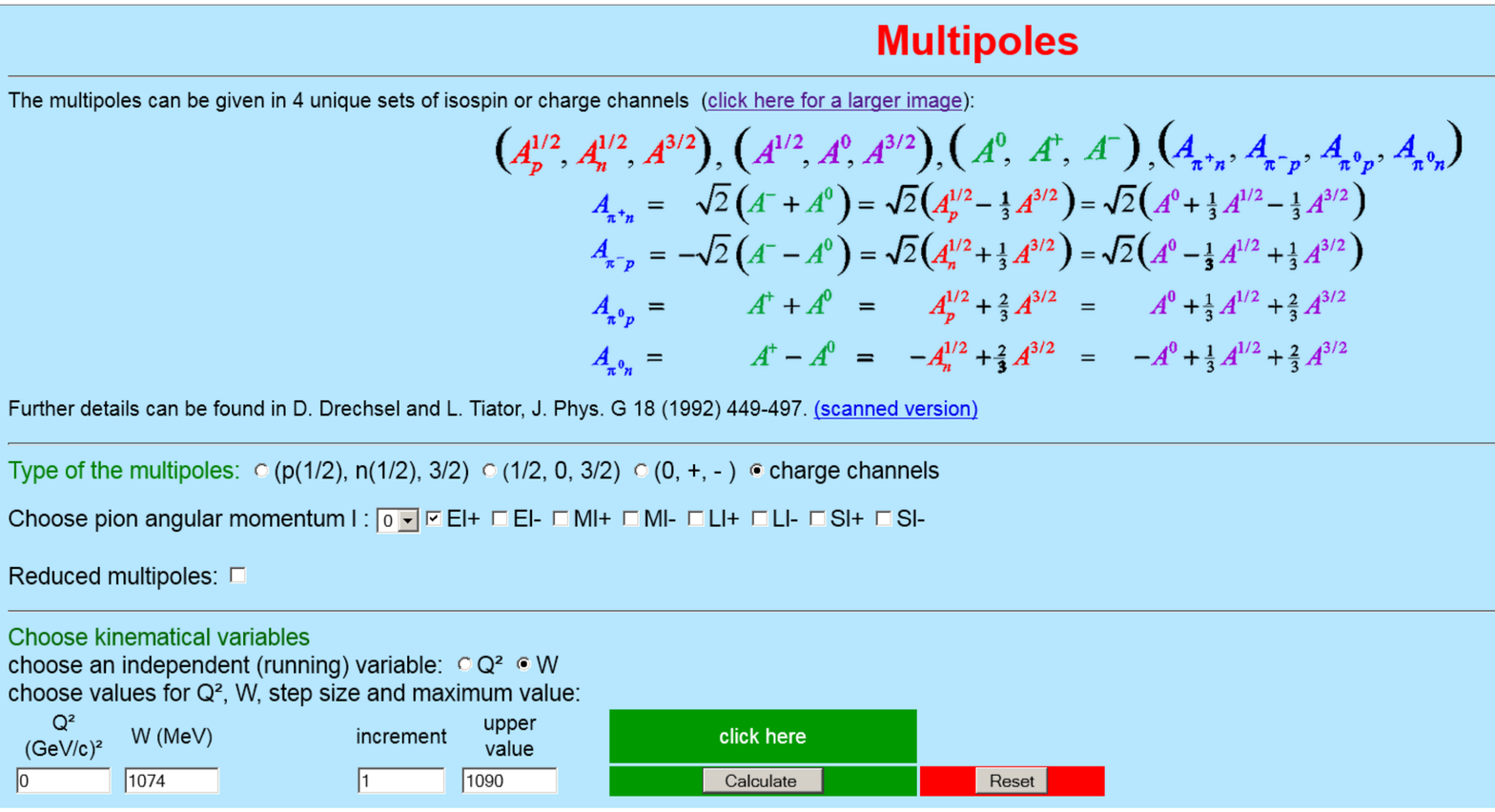}
\caption{\label{fig:Multipoles} Settings to calculate the electric
dipole amplitude $E_{0+}$ for the physical channels.}
\end{center}
\end{figure}

\begin{figure}[ht]
\begin{center}
\includegraphics[width=\textwidth]{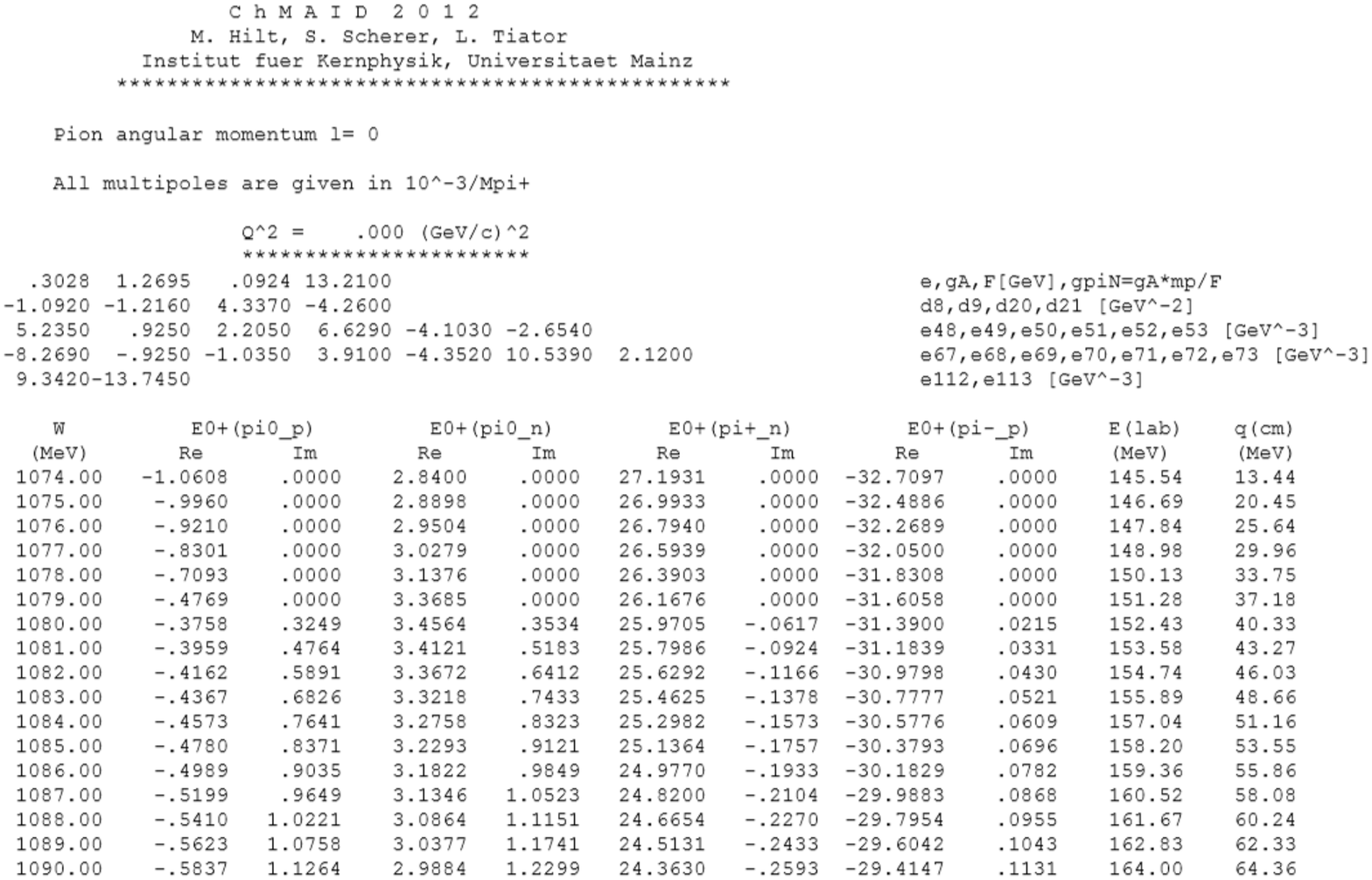}
\caption{\label{fig:Output} Output for the electric
dipole amplitude $E_{0+}$ for the physical channels.}
\end{center}
\end{figure}

\begin{figure}[ht]
\begin{center}
\includegraphics[width=0.8\textwidth]{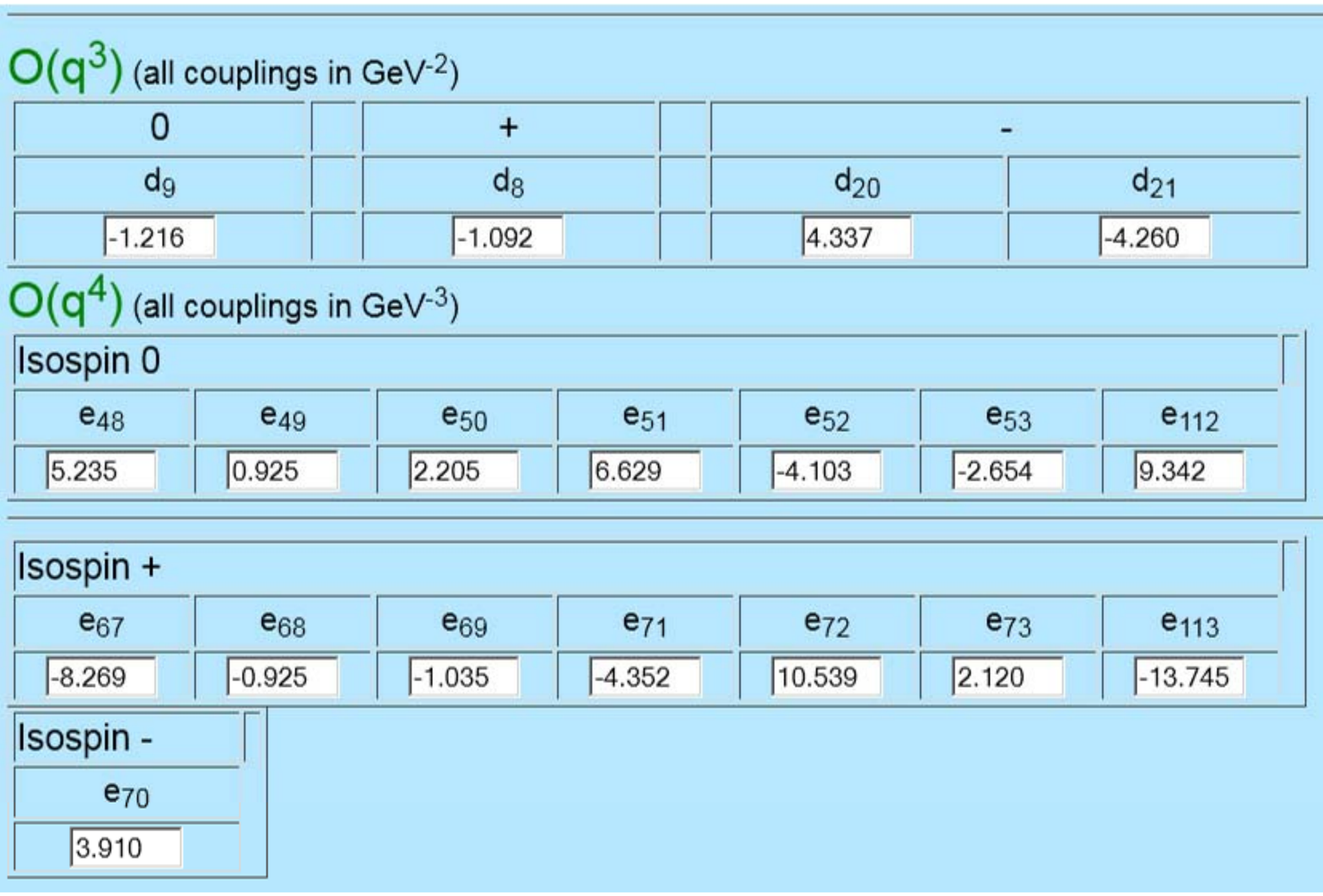}
\caption{\label{fig:Change_of_parameters} The LECs of the contact interactions can be
modified by the user.}
\end{center}
\end{figure}

   As a specific example, Fig.~\ref{fig:Multipoles} shows the settings to calculate the electric
dipole amplitude $E_{0+}$ for the physical channels at the real-photon point as a function
of the total cm energy $W$.
   The corresponding output is shown in Fig.~\ref{fig:Output}.
   The LECs of the contact interactions can be modified by the user
(see Fig.~\ref{fig:Change_of_parameters}).
   The default settings originate from our fit to the available data
(as at year 2013, see Ref.~\cite{Hilt:2013coa}).

   Of course,  $\chi$MAID has a limited range of applicability.
   First of all, ChPT without additional dynamical degrees of freedom restricts the
energy region, where our results can be applied.
   In the case of neutral pion photoproduction (see Ref.~\cite{Hilt:2013uf}) one can clearly
see that for energies above $E_\gamma^\textnormal{lab}\approx 170$ MeV the theory starts to deviate
from experimental data.
   The inclusion of the Delta resonance at ${\cal O}(q^3)$ has recently been discussed in
Refs.~\cite{Blin:2014rpa,Blin:2015lpa}.
   In the case of the charged channels the range of applicability is larger, but some observables are
quite sensitive to the cutoff of multipoles, as the pion pole term is important at small angles.
   As an estimate, for $W>1160$ MeV the difference between our full amplitude and the approximation
up to and including $G$ waves becomes visible.

\section{Results and Conclusions}
\label{sec:results}

\begin{figure}[ht]
    \centering
        \includegraphics[width=0.70\textwidth]{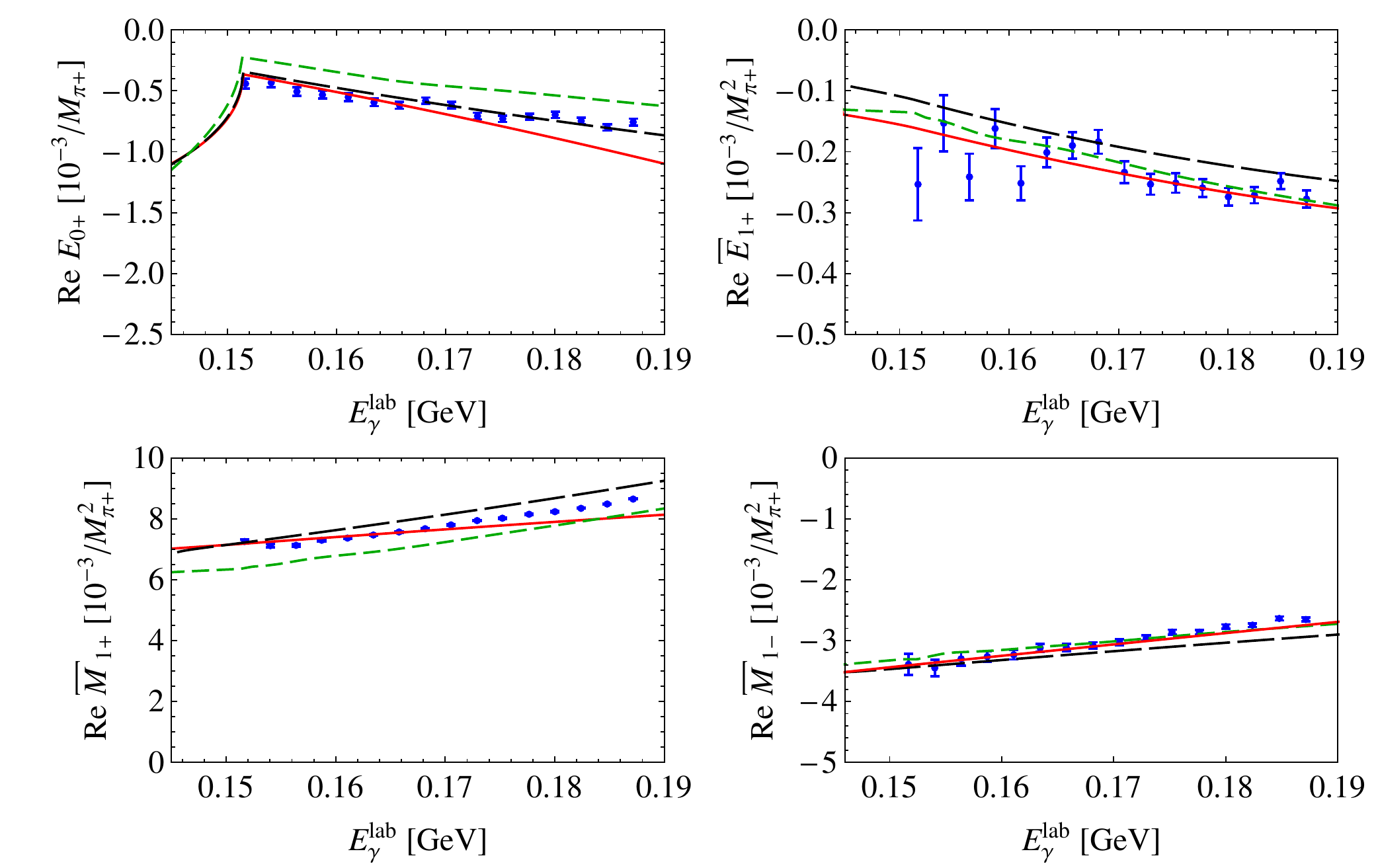}
    \caption{$S$- and reduced $P$-wave multipoles for $\gamma+p\rightarrow p+\pi^0$.
    The solid (red) curves show our RChPT calculations at ${\cal O}(q^4)$.
    The short-dashed (green) and long-dashed (black) curves are
    the predictions of the DMT model \cite{Kamalov:2001qg} and the GL model
    \cite{Gasparyan:2010xz}, respectively.
    The data are from Ref.\ \cite{Hornidge:2012ca}.}
    \label{fig:multipolesexportphyspi0p}
\end{figure}

\begin{figure}[ht]
    \centering
        \includegraphics[width=0.70\textwidth]{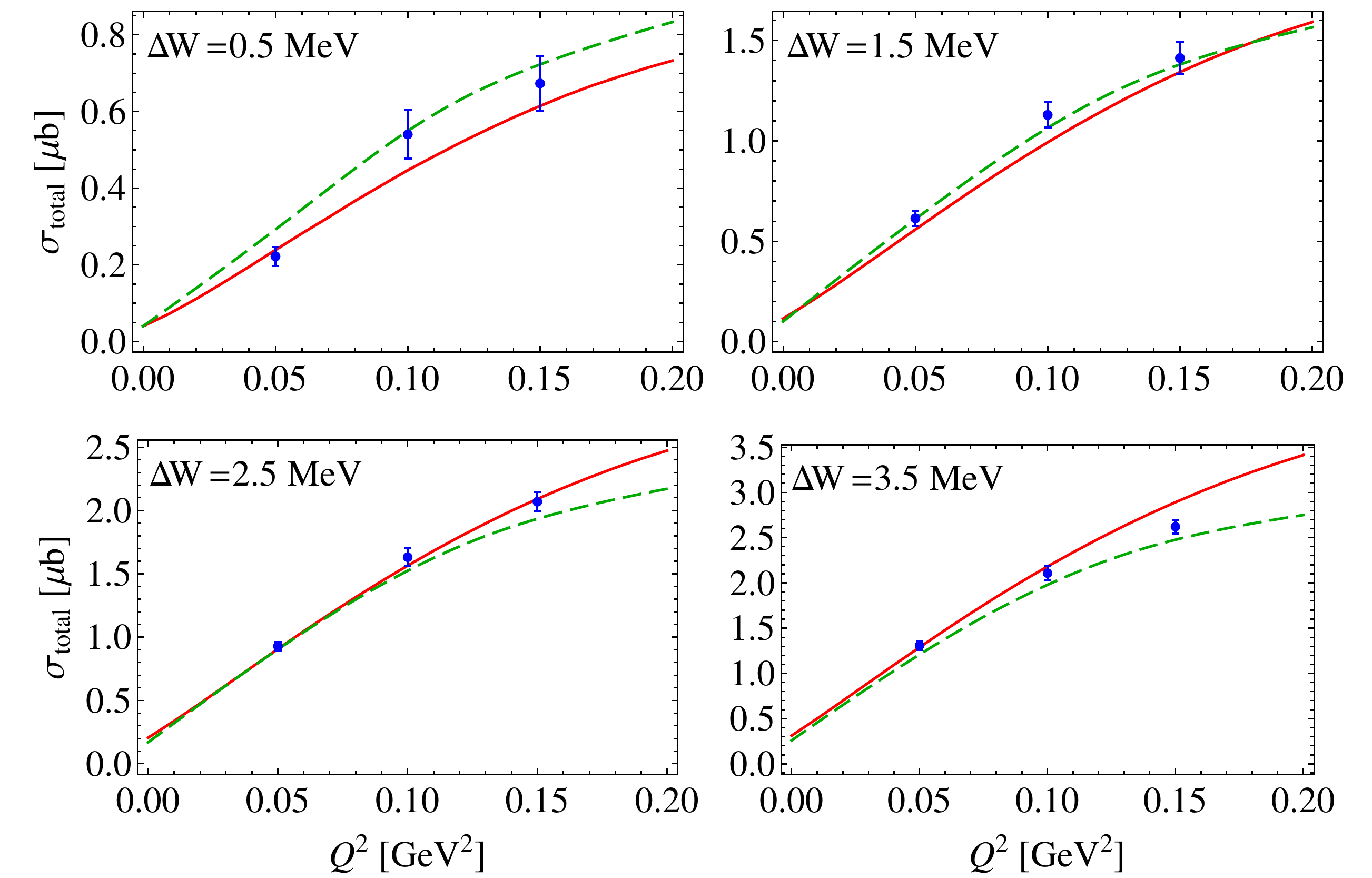}
    \caption{Total cross sections in $\mu$b as a function of $Q^2$ for different cm energies
above threshold $\Delta W$ in MeV. The solid (red) curves show our RChPT calculations at ${\cal O}(q^4)$.
    The short-dashed (green) curves are the predictions of the DMT model \cite{Kamalov:2001qg}.
The data are from Refs.\ \cite{Merkel:20092011}.}
    \label{fig:totcspi0electro}
\end{figure}

   In the following, we present two selected results generated with the chiral MAID
(see Ref.~\cite{Hilt:2013coa} for a complete discussion).
   First,  in Fig.\ \ref{fig:multipolesexportphyspi0p} we show the real parts of the $S$ and $P$ waves
of $\gamma+p\to p+\pi^0$ together with single-energy fits of Ref.\ \cite{Hornidge:2012ca}.
   For comparison, we also show the
predictions of the Dubna-Mainz-Taipei (DMT) model \cite{Kamalov:2001qg} and the covariant, unitary,
chiral approach of Gasparyan and Lutz (GL) \cite{Gasparyan:2010xz}. The multipole $E_{0+}$ agrees
nicely with the data in the fitted energy range. The reduced $P$ waves
$\overline{E}_{1+}=E_{1+}/q_\pi$ and $\overline{M}_{1-}=M_{1-}/q_\pi$ with the pion momentum
$q_\pi$ in the cm frame agree for even higher energies with the single energy fits. The largest
deviation can be seen in $\overline{M}_{1+}$. This multipole is related to the $\Delta$ resonance
and the rising of the data above 170 MeV can be traced back to the influence of this resonance. As
we did not include the $\Delta$ explicitly, this calculation is not able to fully describe its
impact on the multipole.
   For electroproduction, $\gamma^\ast+p\to p+\pi^0$, in Fig.\ \ref{fig:totcspi0electro}, we show the total cross section
$\sigma_{total}=\sigma_T+\epsilon\sigma_L$ in the threshold region together with the experimental
data \cite{Merkel:20092011}.

   In summary we have shown for the first time a chiral perturbation theory approach that can
consistently describe all pion photo- and electroproduction processes in the threshold region equally well.
   By performing fits to the available experimental data, we determined all 19 LECs of the contact graphs
at ${\cal O}(q^3)$ and ${\cal O}(q^4)$ (see Table \ref{tab:lecvalues}).
   Our relativistic chiral perturbation theory calculation is also available online within the MAID
project as chiral MAID under http://www.kph.uni-mainz.de/MAID/.
   It is clear that new experiments will lead to different estimates for the LECs
\cite{Chirapatpimol:2015ftl,Friscic:2015tga}.
   For that reason, we included in $\chi$MAID the possibility of changing the LECs arbitrarily.
   This will help to further study the range of validity and applicability of ChPT in the future.

\begin{table}[htbp]
\centering
\caption{Numerical values of all LECs of pion photo- and electroproduction.
The $\ast$ indicates constants that appear in electroproduction, only.
If possible, the errors were estimated using the bootstrap method (see Ref.~\cite{Hilt:2013coa} for details).
   In the case of the electroproduction LECs $e_{52}$, $e_{53}$, $e_{72}$, and $e_{73}$ we can only give
errors for $\tilde{e}_{52}=e_{52}+e_{72}=6.4\pm 0.7$ and $\tilde{e}_{53}=e_{53}+e_{73}=-0.5\pm 0.2$.}
 \label{tab:lecvalues}
        \begin{tabular}{ccc}
\hline
\hline
  Isospin channel   & LEC  &      Value       \\
\hline
 0 & $d_9\ [\textnormal{GeV}^{-2}]$               &       $-1.22\pm0.12$            \\
 0 & $e_{48}\ [\textnormal{GeV}^{-3}]$            &       $5.2\pm1.4$            \\
 0 & $e_{49}\ [\textnormal{GeV}^{-3}]$            &       $0.9\pm2.6$            \\
 0 & $e_{50}\ [\textnormal{GeV}^{-3}]$            &       $2.2\pm0.8$            \\
 0 & $e_{51}\ [\textnormal{GeV}^{-3}]$            &       $6.6\pm3.6$            \\
 0 & $e_{52}^*\ [\textnormal{GeV}^{-3}]$          &       $-4.1$            \\
 0 & $e_{53}^*\ [\textnormal{GeV}^{-3}]$          &       $-2.7$            \\
 0 & $e_{112}\ [\textnormal{GeV}^{-3}]$           &       $9.3\pm1.6$            \\
\hline
 + & $d_8\ [\textnormal{GeV}^{-2}]$               &       $-1.09\pm0.12$            \\
 + & $e_{67}\ [\textnormal{GeV}^{-3}]$            &       $-8.3\pm1.5$            \\
 + & $e_{68}\ [\textnormal{GeV}^{-3}]$            &       $-0.9\pm2.6$            \\
 + & $e_{69}\ [\textnormal{GeV}^{-3}]$            &       $-1.0\pm2.2$            \\
 + & $e_{71}\ [\textnormal{GeV}^{-3}]$            &       $-4.4\pm3.7$            \\
 + & $e_{72}^*\ [\textnormal{GeV}^{-3}]$          &       $10.5$            \\
 + & $e_{73}^*\ [\textnormal{GeV}^{-3}]$          &       $2.1$            \\
 + & $e_{113}\ [\textnormal{GeV}^{-3}]$           &       $-13.7\pm2.6$            \\
\hline
 $-$ & $d_{20}\ [\textnormal{GeV}^{-2}]$            &       $4.34\pm0.08$            \\
 $-$ & $d_{21}\ [\textnormal{GeV}^{-2}]$            &       $-3.1\pm0.1$            \\
 $-$ & $e_{70}\ [\textnormal{GeV}^{-3}]$            &       $3.9\pm0.3$            \\
\hline
\hline
\end{tabular}

\end{table}

This work was supported by the Deutsche Forschungsgemeinschaft (SFB 443 and 1044).

\end{document}